\documentstyle[psfig]{elsart}

\newcommand{\kms}{\mbox{km s$^{-1}$}}

\newcommand{\msun}{\mbox{$M_{\odot}$}}

\newcommand{\kmsmpc}{\mbox{km s$^{-1}$ Mpc$^{-1}$}}

\def\la{\mathrel{\hbox{\rlap{\hbox{\lower4pt\hbox{$\sim$}}}\hbox{$<$}}}}
\def\ga{\mathrel{\hbox{\rlap{\hbox{\lower4pt\hbox{$\sim$}}}\hbox{$>$}}}}

\def\LCDM{$\Lambda$CDM}

\begin{document}
\begin{frontmatter}
\title{The nature of high-redshift galaxies\thanksref{pub}}
\author[ad1]{Joel R. Primack},
\author[ad2]{Rachel S. Somerville,}
\author[ad3]{S. M. Faber,}
\author[ad1]{Risa H. Wechsler}
\thanks[pub]{To appear in the Proceedings of the 3rd International
Symposium on Sources and Detection of Dark Matter in the Universe 
(DM98), Feb. 1998, ed. D. Cline.}
\address[ad1]{Physics Department, University of California, \cty Santa
Cruz, CA 95060 \cny USA}
\address[ad2]{Racah Institute of Physics, The Hebrew University,
\cty Jerusalem 91904, \cny Israel}
\address[ad3]{UCO/Lick Observatory, University of California, \cty
Santa Cruz, CA 95060 \cny USA}

\begin{abstract}
Using semi-analytic models of galaxy formation, we investigate the
properties of $z\sim3$ galaxies and compare them with the observed
population of Lyman-break galaxies (LBGs).  In addition to the usual
quiescent mode of star formation, we introduce a physical model for
starbursts triggered by galaxy-galaxy interactions. We find that with
the merger rate that arises naturally in the CDM-based merging
hierarchy, a significant fraction of bright galaxies identified at
high redshift ($z \ga 2$) are likely to be low-mass, bursting
satellite galaxies. The abundance of LBGs as a function of redshift
and the luminosity function of LBGs both appear to be in better
agreement with the data when the starburst mode is included,
especially when the effects of dust are considered. The objects that
we identify as LBGs have observable properties including low velocity
dispersions that are in good agreement with the available data. In
this ``Bursting Satellite'' scenario, quiescent star formation at
$z\ga2$ is relatively inefficient and most of the observed LBGs are
starbursts triggered by satellite mergers within massive halos. In
high-resolution N-body simulations, we find that the most massive dark
matter halos cluster at redshift $z\sim 3$ much as the 
LBGs are observed to do. This is true for both the
$\Omega=1$ CHDM model and low-$\Omega$ \LCDM\ and OCDM models, all of
which have fluctuation power spectra $P(k)$ consistent with the
distribution of low-redshift galaxies.  The Bursting Satellite
scenario can resolve the apparent paradox of LBGs 
that cluster like massive dark matter halos but have narrow
linewidths and small stellar masses.
\end{abstract}

\end{frontmatter}

\section{Introduction}

Our window onto the high redshift universe ($z\ga 2$) has been expanded
tremendously by the ``Lyman-break'' photometric selection technique
developed by Steidel and collaborators \cite{steidel93,steidel96a}. Similar
techniques were exploited by Madau et al. 
\cite{madau96} to identify high-redshift
candidates in the Hubble Deep Field (HDF). Extensive
spectroscopic follow-up work at the Keck telescope has verified the
accuracy of the photometric selection technique
\cite{steidel96b,lowenthal97}. The morphologies and sizes of these objects
can be studied using the HDF sample \cite{giavalisco96,lowenthal97}, and
their clustering properties can be studied using the growing sample of
hundreds of Lyman-break galaxies (LBGs) with spectroscopic redshifts
\cite{steidelspike,adelberger98}.

One interpretation of LBGs \cite{steidel96a,giavalisco96,steidelspike}
is that they are located in the centers of massive dark matter halos ($M
\sim 10^{12} \msun$) and have been forming stars at a moderate rate over a
fairly long time-scale ($\ga 1$ Gyr). This ``Massive Progenitor'' scenario
supposes that the galaxies identified as LBGs at $z\sim3$ are the
progenitors of the centers of today's massive luminous ellipticals and
spheroids, around which the outer parts later accrete. This viewpoint
has been supported by semianalytic modeling \cite{bcfl}. But even though
the observed clustering of LBGs is very similar to that of massive dark
matter halos in N-body simulations \cite{jingsuto,bagla,wechsler}, it does
not necessarily follow that the Massive Progenitor scenario is correct.

Here we consider the viability of an alternative interpretation of the
observations \cite{sfp,spf98}. There is clear observational evidence that
galaxy-galaxy interactions trigger ``starbursts'', a mode of star formation
with a sharply increased efficiency over a relatively short timescale.
There is also observational evidence that galaxy-galaxy interactions and
starbursts are more common at high redshift than they are today
\cite{guzman97}. The similarity between the appearance of the spectra of
the LBGs and local starburst galaxies has been noted
\cite{lowenthal97,pettini97}. It seems likely that at least some of the
observed Lyman-break galaxies are relatively low-mass ($\sim 10^{9}-10^{10}
\msun$) objects in the process of an intense starburst, plausibly triggered
by galaxy encounters. If a significant fraction of the objects are of this
nature, this would have far-reaching implications for the interpretation of
the observations. In this alternative ``Bursting Satellite'' scenario, we
predict that the population of bright galaxies at high redshift would still
be found in massive dark matter halos, but they would not necessarily
evolve into the centers of elliptical or spiral galaxies; instead, many of
them may form the low-metallicity Population II stellar halos that surround
bright galaxies at low redshift \cite{trager,searle-zinn}, and they could
also be a major source of enriched gas in massive dark matter halos. The
objects that we identify as LBGs in our models have star formation rates,
half-light radii, I$-$K colors, and velocity dispersions that are in good
agreement with the available data. In Ref. \cite{spf98} we also investigate
global quantities such as the star formation rate density and cold gas and
metal content of the universe as a function of $z$.

\section{Clustering of LBGs}

The Bursting Satellite scenario predicts that LBGs are mainly in massive
dark matter halos. That LBGs are associated with massive dark matter halos
is a natural interpretation of the strong clustering in redshift exhibited
by the LBGs \cite{steidelspike,adelberger98}. Such halos at high redshift
are expected to cluster much more strongly than the underlying dark matter,
since they will be located mainly where sheets and filaments of dark matter
intersect.

In order to investigate quantitatively whether this model agrees with the
observed redshift clustering of the LBGs, we \cite{wechsler} assumed that
the LBGs are associated with the most massive dark matter halos at $z\sim3$
in a suite of high-resolution cosmological simulations
\cite{gross97,grossea98} of several of the most popular CDM-variant
cosmologies: standard cold dark matter (SCDM) with $\Omega=1$, $h=0.5$, and
$\sigma_8=0.67$, and four COBE-normalized models: CDM with $\Omega=1$,
$h=0.5$, and $\sigma_8=1.3$; cold plus hot dark matter (CHDM) with
$\Omega=1$, $h=0.5$, and $\Omega_\nu=0.2$ in $N_\nu=2$ species of light
neutrinos; open CDM (OCDM) with $\Omega=0.5$ and $h=0.6$; and a flat CDM
cosmology (\LCDM) with $\Omega=0.4$, $\Omega_\Lambda \equiv
\Lambda/(3H_0^2) = 1-\Omega_0$, and $h=0.6$. These simulations were run
in 75 $h^{-1}$ Mpc boxes with 57 million cold particles and a dynamic range
in force resolution of $\sim10^3$; they thus had adequate resolution to
identify all dark matter halos with a comoving number density at
$z\sim3$ higher than that of 
the observed LBGs. Since redshifts were measured
spectroscopically \cite{steidelspike} for only about 40\% of the
photometric LBG candidates, we chose a mass threshold for dark matter halos
in the relevant interval in redshift in each simulation such that the
comoving number density of the halos would be equal to that of the LBG
candidates, and then we sampled 40\% of these at random. When we compared
the clustering statistics of these halos to the observed distribution of
the redshifts corrected for selection, we found that the CHDM, OCDM, and
\LCDM\ simulations predicted that pencil beams as long and wide as the
first published one \cite{steidelspike} would have a ``spike'' in the
redshift distribution at $z\sim3$
as high as the one actually observed approximately
1/3 of the time, which appears consistent with the additional statistics
now available \cite{adelberger98}. However, the CDM model normalized either
to clusters or to COBE had a spike as high as this less than 10\% of the
time. The distribution on the sky of these massive halos was very similar
to that of the observed LBGs in all the simulations; i.e., they form
extended structures approximately 10 $h^{-1}$ Mpc across. So the spikes are
not yet clusters at $z\sim3$, but we find that they evolve into at least
Virgo-size clusters by $z=0$ (cf. \cite{governato}).

The autocorrelation function $\xi(r)$ of the massive halos
that we identify with LBGs in each simulation \cite{wechsler} has a
power-law index of about -1.5, somewhat shallower than the
-1.8 to -2.0 observed \cite{giavalisco98}, and the correlation lengths that
we calculate are a little larger than observed. But we expect that the
agreement will improve when we take into account the higher autocorrelation
of the more massive halos and the higher probability that these will host
LBGs. We are presently doing this calculation combining our simulations
with the semi-analytic models discussed in the next section.

\section{Semi-analytic models of galaxy formation}

Semi-analytic techniques allow one to model the formation and evolution of
galaxies in a hierarchical framework, including the effects of gas cooling,
star formation, supernova feedback, galaxy-galaxy merging, and the
evolution of stellar populations. The semi-analytic models used here are
described in detail in \cite{mythesis,sp98,spf98}. These models are in
reasonably good agreement with a broad range of local galaxy observations,
including the Tully-Fisher relation, the B-band luminosity function, cold
gas contents, metallicities, and colors. Our basic approach is similar in
spirit to the models originally developed by the Munich \cite{kwg93} and
Durham \cite{cafnz94} groups, and subsequently elaborated by these groups
in numerous other papers (reviewed in \cite{mythesis,sp98}). We have
reproduced much of the work of these groups, and improved on it in modeling
the low-redshift universe in three main ways: (1) correcting the local
Tully-Fisher normalization; (2) including extinction due to dust, which
is crucial to get correctly both the Tully-Fisher relation (always
corrected for extinction) and the luminosity function (not corrected for
extinction); and (3) developing an improved disk-halo treatment of the
energy and metals in supernova ejecta.

The framework of the semi-analytic approach is the ``merging history'' of a
dark matter halo of a given mass, identified at $z=0$ or any other redshift
of interest. We construct Monte-Carlo realizations of the ``merger trees''
using an improved method \cite{sk98}. Each branch in the tree represents a
halo merging event. When a halo collapses or merges with a larger halo, we
assume that the associated gas is shock-heated to the virial temperature of
the new halo. This gas then radiates energy and cools. The cooling rate
depends on the density, metallicity, and temperature of the gas. Cold gas
is turned into stars using a simple recipe with the stellar masses assumed
to follow the standard Salpeter IMF, and supernova energy reheats the cold
gas according to another recipe. The free parameters are set by requiring
an average fiducial ``reference galaxy'' (the central galaxy in a halo with
a circular velocity of $220 \,\kms$) to have an I-band magnitude $M_{I}
-5\log h = -21.7$ (this requirement fixes the zero-point of the I-band
Tully-Fisher relation to agree with observations), a cold gas mass
$m_{\rm cold} = 1.25\times 10^{10} h^{-2} \msun$, and a stellar metallicity
of about solar. The star formation and feedback processes are some of the
most uncertain elements of these models, and indeed of any attempt to model
galaxy formation. As in our investigation of local galaxy properties
\cite{sp98}, we have considered several different combinations of recipes
for star formation and supernova feedback (sf/fb) and also several
cosmologies \cite{spf98}, but here we will report high-redshift results
only for a single choice of sf recipe (SFR-D) and cosmology (SCDM).

\subsection{Modeling starbursts}

Previous semi-analytic models have not systematically investigated the
importance of a bursting mode of star formation, particularly its effect on the
interpretation of the observations of high-redshift galaxies. We start with the
ansatz that galaxy-galaxy mergers trigger starbursts. This premise has
considerable observational support and is also supported by N-body simulations
with gas dynamics and star formation \cite{mihos94,mihos96}.

In our models, galaxies that are within the same large halo may merge
according to two different processes. Satellite galaxies lose energy and
spiral in to the center of the halo on a {\it dynamical friction}
time-scale. In addition, satellite galaxies orbiting within the same halo
may merge with one another according to a modified {\it mean free path}
time-scale. Our modeling of the latter process is based on the scaling
formula derived in Ref. \cite{makino-hut} to describe the results of
dissipationless N-body simulations in which galaxy-galaxy encounters and
mergers  were simulated, covering a large region of parameter space.

When any two galaxies merge, the ``burst'' mode of star formation is turned
on, with the star formation rate during the burst modeled as 
a Gaussian function of
the time. The burst model has two adjustable parameters, the time-scale of
the burst and the efficiency of the burst (the fraction of the cold gas
reservoir of both galaxies combined that is turned into stars over the
entire duration of the burst). The timescale and efficiency parameters that
we use are based on the simulations \cite{mihos94,mihos96} mentioned above,
treating major ($m_{\rm smaller}/m_{\rm larger} > f_{\rm bulge} \sim 0.3$)
and minor mergers separately. The quiescent mode of star formation
continues as well. Details are given in \cite{spf98}.

\begin{figure}
\centerline{\psfig{file=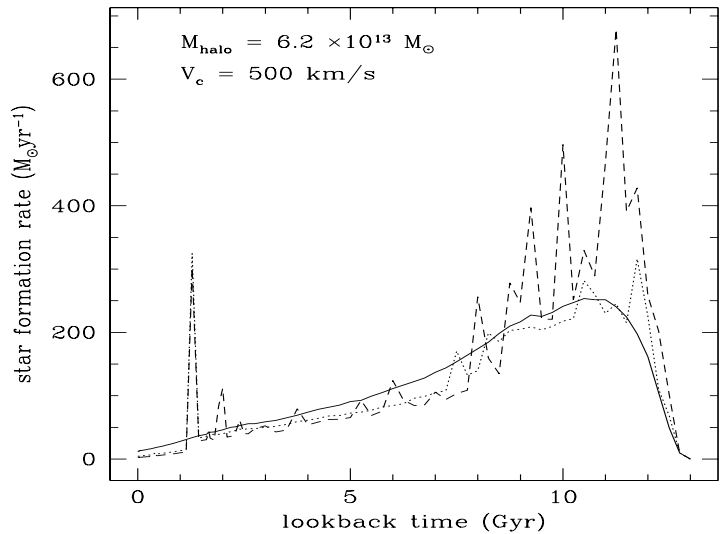,height=7.5truecm,width=10truecm}}
\caption{The total star formation rate for all galaxies that end up within a
typical halo with $V_c = 500\, \kms$ at $z=0$. The solid line shows a model (1)
with no starbursts, the dotted line shows a model (2) with mergers only
between satellite and central galaxies and bursts only in major mergers,
and the dashed line shows a model (3) with satellite-satellite and
satellite-central mergers and bursts in major and minor mergers. The
lookback time is computed for $\Omega=1$ and $H_0=50\,\kmsmpc$. Note that
in model (3) the peaks, representing starbursts, occur primarily at
lookback times of 8 to 12 Gyr (redshifts $z \sim 1-5)$. }
\label{fig:sfhist_group}
\end{figure}
Fig.~\ref{fig:sfhist_group} shows the total star formation rate for
all the galaxies in a large group halo ($V_c = 500\,\kms$ at
$z=0$). The star formation rate is shown in models with: (1) no
starbursts (quiescent star formation only), (2) bursts in major
mergers only and satellite galaxies only allowed to merge with the
central galaxy on a dynamical friction time-scale (no
satellite-satellite mergers) and (3) satellite-central and
satellite-satellite mergers, and bursts in both major and minor
mergers. The star formation rate at high redshift is considerably
amplified in model (3) compared to models (1) and (2), illustrating
that neglecting satellite-satellite mergers and bursts in minor
mergers will considerably underestimate the importance of starbursts
at high redshift. The ``burst'' models discussed in the remainder of
this paper correspond to the maximal burst scenario, model (3) above.
(The models of \cite{bcfl} are similar to (2).)

\subsection{Comoving number density of LBGs}
\begin{figure}
\centerline{\psfig{file=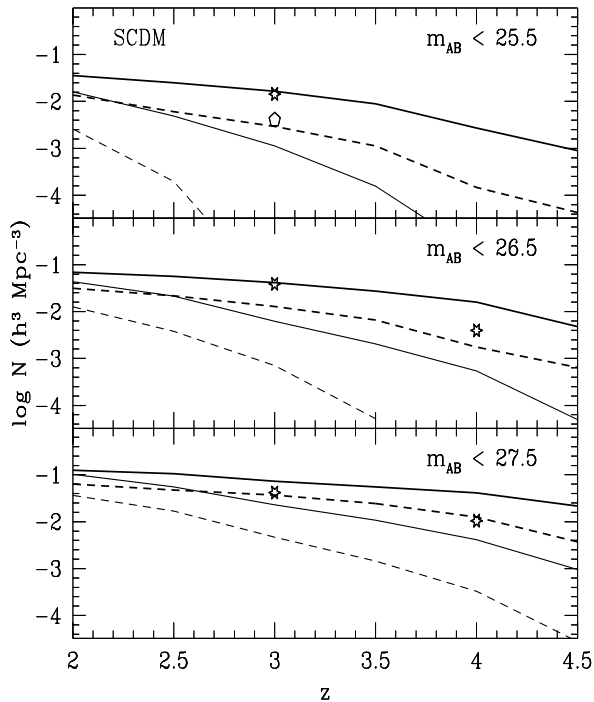,height=10truecm,width=8truecm}}
\caption{The comoving number density of galaxies brighter than $m_{\rm lim}$,
where $m_{\rm lim}=25.5$ (top panel), 26.5 (middle panel), or 27.5 (bottom
panel). The hexagon indicates the comoving number density of LBGs with
spectroscopic redshifts from the ground-based sample of
\protect\cite{giavalisco98} and the stars indicate the number
density of U ($z\sim3$) and B ($z\sim4$) drop-outs in the HDF
\protect\cite{madau96}. Bold solid lines show the comoving number density
of galaxies in the models with starbursts; light solid lines show the
results of the no-burst models. Dashed lines show the result of reducing
the flux of each galaxy by a factor of three to estimate the effects of
dust extinction \protect\cite{pettini97}. }
\label{fig:counts_scdm}
\end{figure}
The first question is whether the models reproduce the observed number
densities of objects at high $z$. Fig.~\ref{fig:counts_scdm} shows the
comoving number density of galaxies brighter than a fixed magnitude limit
as a function of redshift, over the redshift range probed by the observed
LBGs. We show this function using three values for the magnitude limit: the
top panel shows the abundance of bright LBGs ($m < 25.5$), the middle panel
shows the abundance of galaxies brighter than $26.5$, and the bottom panel
shows galaxies brighter than $27.5$. We have calculated the comoving number
density for the ground-based sample of LBGs with spectroscopic redshifts
using data from Ref. \cite{giavalisco98}, and the comoving number density
of LBGs at $z\sim3$ and $z\sim4$ from the HDF using the list of $U$ and $B$
drop-outs from Ref. \cite{madau96} (see \cite{spf98} for details). We note
that the number density of LBGs in the HDF is considerably higher than for
the ground-based sample. This may be an indication that the ground based
sample is missing a substantial number of objects, or it may indicate that
the small HDF volume probed by the $U$ and $B$ dropouts is an unusually
overdense region.

Fig.~\ref{fig:counts_scdm} shows that models without bursts underpredict
the number density of galaxies, especially at the brightest magnitude limit
$m=25.5$. Burst models reproduce or exceed the observed number
densities of LBGs when dust extinction is neglected. The inclusion of
starbursts causes a bigger change in the comoving number density of LBGs at
higher redshifts and at brighter magnitude limits; i.e., the number
densities in the burst models tend to have flatter dependences on both
redshift and magnitude. By a redshift of $z\la2$, including starbursts has
little effect on the number counts. The galaxy-galaxy merger rate is larger
at high redshift because the halos are denser, and the starbursts are more
dramatic because these galaxies are relatively gas rich.

The inclusion of dust is an important correction. The observed colors of
the LBGs, as well as comparison of the UV to H$\beta$ fluxes, indicate that
there is almost certainly \emph{some} dust in these galaxies
\cite{pettini97,dickinson98,calzetti97b}. However, the amount of dust and
the resulting extinction are quite uncertain. These depend on the
metallicity and age of the galaxy, the geometry and ``clumpiness'' of the
dust, and the wavelength dependence of the attenuation law. The correction
factors for the UV rest frame luminosity suggested in 
\cite{pettini97,dickinson98,calzetti97b}
range from $\sim2$ to $\sim7$. More dramatic corrections, as large as a
factor of $\sim 15$, have been suggested \cite{meurer,sawickiyee98}.

Our estimates of the effect of dust in Fig.~\ref{fig:counts_scdm}
simply decrease the luminosity of each galaxy by a factor of
three. However, according to any physical dust model, a uniform
correction by a fixed factor is probably unrealistic. If dust traces
metal production (and hence star formation activity), more
intrinsically luminous galaxies will be more heavily extinguished. It
seems unavoidable that this will further increase the deficit of
bright galaxies in the no-burst models seen in
Fig.~\ref{fig:counts_scdm}. However, if most of the bright galaxies
are starbursting objects, as in the burst models, the situation is
less clear.  Observations \cite{calzetti94,calzetti97a} indicate that
the wavelength dependence of the attenuation due to dust is ``greyer''
(less steep) in the UV for local starburst galaxies than a Galactic or
SMC-type extinction curve. Powerful starbursts could blow holes in the
dust, especially in small objects, perhaps ejecting the dust (along
with metals) out of the galaxy. On the other hand, regions of active
star formation may be completely enshrouded in dust, leading to even
stronger extinction. In any case, models without starbursts appear to
have no hope of reproducing the observed abundance of bright LBGs with
just the conservative factor of three correction for dust included in
Fig.~\ref{fig:counts_scdm}. Even our models including starbursts do
not reproduce the observed abundance of the brightest HDF LBGs with
this dust correction. Since the light observed in the visible was
emitted as ultraviolet in the rest frame, changing from the assumed
Salpeter IMF to one with more high-mass stars can significantly
increase the predicted abundance of bright LBGs. However, even such
top-heavy IMFs will probably not be sufficient to save the no-burst
models, which also predict a LBG luminosity function that is too steep
compared to observations \cite{spf98}.

\subsection{Line-widths, ages, and masses}
The velocity dispersions of observed LBGs can be estimated based on the
widths of stellar emission lines such as H$\beta$ or O[III]. Emission lines
have been detected for a few of the brightest LBGs from the ground-based
sample. The velocity dispersions $\sigma$
derived from the observed linewidths are
$\sigma=80-90 \,\kms$ for four objects, $50 \,\kms$ for one object, and
$150\,\kms$ for one object \cite{pettinilw}. 
These values agree well with the burst models, for which the
probability distribution for $\sigma$ of the stars
in a disk geometry peaks at  $\sim 80 \, \kms$, but are in strong
disagreement with the no-burst models, which peak at $\sigma\sim180 \,
\kms$.  

These measurements may be affected by several biases, which remain to
be unravelled.  First, the observed data refer to bright LBGs, which
may have systematically higher linewidths.  This effect would increase
the discrepancy with no-burst models mentioned above.  The modelling
of the velocity dispersion at the small radii probed by the
observations (approximately the half-light radius) is also uncertain
because of the uncertain morphology of the observed LBGs (disk
vs. spheroid).  This should be studied with high-resolution narrowband
imaging.  Still, overall the present $\sigma$s suggest small galaxies
whose brightnesses are being amplified temporarily by starbursts.

The ages and stellar masses of the burst models \cite{spf98} also
agree well with recent estimates based on observed SEDs including IR
photometry.  LBG colors are well fit by young ($<0.1$ Gyr) stellar
populations with moderate amounts of dust \cite{sawickiyee98}.  These
young ages imply that stellar masses are also low ($\sim 10^9
M_\odot$).  The LBGs in our no-burst models (and also Ref. \cite{bcfl}) are
systematically older and more massive than the data indicate.  More
photometry and 
spectra of LBGs will help to clarify whether our models including
``Bursting Satellites'' adequately describe the properties of the
LBGs.

\section*{Acknowledgment}
JRP acknowledges support from a NASA ATP grant at UCSC; RSS, a
GAANN Fellowship at UCSC and a University Fellowship from The Hebrew
University; and RHW, a Cota-Robles Fellowship at UCSC.

\end{document}